\documentclass[11pt]{article}
\usepackage{latexsym}
\usepackage{amssymb}
\usepackage[dvips]{graphicx}
\usepackage{epsfig}
\usepackage{rotating}

\usepackage{amsfonts}
\usepackage{amsmath}
\newlength{\Taille}

\input xy
\xyoption{all}

\newcommand{\flechebas}[1]{
  \settoheight{\unitlength}{\mbox{$#1$}}
  \settowidth{\Taille}{\mbox{~${\scriptstyle #1}$}}
  \addtolength{\unitlength}{4ex}
  \begin{picture}(0,1)
    \put(0,1){\vector(0,-1){1}}
    \put(0,0.5){\makebox(0,0){${\scriptstyle #1}$ \hspace{\the\Taille}}}
  \end{picture}}
\newcommand{\flechehaut}[1]{
  \settoheight{\unitlength}{\mbox{$#1$}}
  \settowidth{\Taille}{\mbox{~${\scriptstyle #1}$}}
  \addtolength{\unitlength}{4ex}
  \begin{picture}(0,1)
    \put(0,0){\vector(0,1){1}}
    \put(0,0.5){\makebox(0,0){\hspace{\the\Taille}${\scriptstyle #1}$ }}
  \end{picture}}
\newcommand{\flechedroite}[1]{
  \settowidth{\unitlength}{\mbox{$#1$}}
  \settoheight{\Taille}{\mbox{${\scriptstyle #1}$}}
  \addtolength{\Taille}{1ex}
  \addtolength{\unitlength}{4ex}
  \raisebox{0.5ex}{
  \begin{picture}(1,0)
    \put(0,0){\vector(1,0){1}}
    \put(0.5,0){\makebox(0,0){${\scriptstyle #1}$ \vspace{\the\Taille}}}
  \end{picture}}}
\newcommand{\flechegauche}[1]{
  \settowidth{\unitlength}{\mbox{$#1$}}
  \settoheight{\Taille}{\mbox{${\scriptstyle #1}$}}
  \addtolength{\Taille}{1ex}
  \addtolength{\unitlength}{4ex}
  \raisebox{0.5ex}{
  \begin{picture}(1,0)
    \put(1,0){\vector(-1,0){1}}
    \put(0.5,0){\makebox(0,0){${\scriptstyle #1}$ \vspace{\the\Taille}}}
  \end{picture}}}

\newtheorem{proposition}{Proposition}[section]

\newtheorem{remark}{Remark}[section]

\begin{document}
\begin{titlepage}
\begin{center}
{\Large \bf {Harmonic oscillator in twisted Moyal plane:
eigenvalue problem and relevant properties}}

Mahouton Norbert Hounkonnou$^{1,\dag}$ and Dine Ousmane
Samary$^{1,*}$

 $^{1}${\em University of Abomey-Calavi,\\
International Chair in Mathematical Physics
and Applications}\\
{\em (ICMPA--UNESCO Chair), 072 B.P. 50  Cotonou, Republic of Benin}\\

E-mails:   $^{\dag}$norbert.hounkonnou@cipma.uac.bj,\\
$^{*}$ousmanesamarydine@yahoo.fr.

\begin{abstract}
The paper reports on a study of a harmonic oscillator  (ho) in the twisted Moyal space,
in a well defined matrix basis, generated by the vector
 fields $X_{a}=e_{a}^{\mu}(x)\partial_{\mu}=(\delta_{a}^{\mu}+\omega_{ab}^{\mu}x^{b})\partial_{\mu}$, which
induce a dynamical star product. The usual multiplication law can be hence reproduced
 in the  $\omega_{ab}^{\mu}$ null limit. The star actions of creation
and annihilation  functions are explicitly computed. The ho states are infinitely
degenerate with energies depending on the coordinate functions.
\end{abstract}


\end{center}
\end{titlepage}

\section{Introduction}
It is generally believed that the picture of spacetime as a manifold M locally modelled on the flat Minkowski
 space should break down at
very short distances of the order of the Planck length $l_{p}=(G\hbar/c^3)^{1/2}$. Limitations in
the possible accuracy of localization of
 spacetime events should in fact be a feature of a quantum theory incorporating gravitation.
The obtaining of a better understanding of physics at short distances and the cure of the problems occuring
 when trying to quantize gravity should lead to change the nature of spacetime
in a fundamental way. This could be realized by implementing the  noncommutativity through the coordinates which satisfy
the commutation relations
$[\hat{x}^{\mu},\hat{x}^{\nu}]=iC^{\mu\nu}(\hat{x})\neq 0$.
In general, the function $C^{\mu\nu}(\hat{x})$ is unknown, but, for physical
 reasons, should  vanish at large distances where we experience
the commutative world and may be determined by experiments \cite{doplicher}
 and \cite{frank}. The $\Theta-$deformation
case which may at very short distances provide a reasonable
approximation for $C^{\mu\nu}(\hat{x})$ is described by the
commutation relation
$[\hat{x}^{\mu},\hat{x}^{\nu}]=i\Theta^{\mu\nu}$.
$\Theta^{\mu\nu}$ is usually chosen in the form
\begin{eqnarray}
\Theta = \left(\begin{array}{llllllll}
0 & \Theta_{1} &  &  &  &  &  &  \\
-\Theta_{1}  &0  &  &  &  &0  &  &  \\
 &  & 0 &\Theta_{2}  &  &  &  &  \\
 &  &- \Theta_{2} & 0 &  &  &  &  \\
 &  &  &  &\vdots & \vdots &  &  \\
 &  &  &  & \vdots & \vdots &  &  \\
 &  &0  &  &  &  & 0 &\Theta_{\frac{D}{2}}  \\
 &  &  &  &  &  & -\Theta_{\frac{D}{2}} & 0
\end{array}\right)
\end{eqnarray}
where $\Theta_{j}\in {\mathbb{R}}$, $j=1,2,\cdots, \frac{D}{2}$,
have dimension of length square, ($[\Theta_j] = [L]^2$), $D$
denoting the spacetime dimension.
The algebra of functions of such noncommuting coordinates can be represented
by the algebra of functions on ordinary spacetime, equipped with a noncommutative $\star-$product.
 For a constant antisymmetric matrix
 $\Theta^{\mu\nu}$, this can be represented by the Groenewold-Moyal product:
\begin{eqnarray}
 (f\star g)(x) = {\rm m}\Big \{e^{i \frac{\Theta^{\rho\sigma}}{2}
 \partial_{\rho}\otimes \partial_{\sigma}} f(x)\otimes g(x)
 \Big \} \quad x\in {\rm I\!\!R}_\Theta^D \quad  \forall f,g\in C^{\infty}({\rm I\!\!R}_\Theta^D)
 \end{eqnarray}
${\rm m}$ is the   ordinary multiplication of functions,
$C^{\infty}({\rm I\!\!R}_\Theta^D)$ - the space of suitable smooth functions on ${\rm I\!\!R}_\Theta^D$
and ${\rm I\!\!R}_\Theta^D$ - the $D-$dimensional Moyal space. This product can be generalized under the form
\begin{eqnarray}
(f\star g)(x)={\rm m}\Big\{e^{i\frac{\Theta^{ab}}{2}X_{a}\otimes X_{b}}f(x)\otimes g(x)\Big\}
\end{eqnarray}
where $X_{a}=e_{a}^{\mu}(x)\partial_{\mu}$ are vector fields. The
commutation relation of coordinates then becomes
$[x^{\mu},x^{\nu}]_{\star}=i\Theta^{ab}e_{a}^{\mu}(x)e_{b}^{\nu}(x)
:=i\widetilde{\Theta}^{\mu\nu}(x)$ engendering a twisted scalar
field theory where $e_{a}^{\mu}$, and hence the $\star-$product
itself, appear dynamical. See \cite{aschieri}-\cite{ hd} for more details. Besides,
the Leibniz rule extends to the commuting fields $X_{a}$ as
follows: \begin{eqnarray}\label{leib} X_{a}(f\star
g)=(X_{a}f)\star g+f\star(X_{a}g).\end{eqnarray}

Recently \cite{aschieri}, a formulation of dynamical noncommutativity,
which allows for a consistent interpretation of position measurement and the solution of
 the  problem of a noncommutative well has been put forward. This work  addresses a study of
a harmonic oscillator properties in the twisted Moyal plane.

The paper is organized as follows. In Section 2, using appropriate matrix basis
and  deforming the issue of a twisted product, we solve  the resulting  eigenvalue problem to find
 the states and the energy spectrum of the harmonic oscillator Hamiltonian. These states are infinitely degenerate.
Some concluding remarks are pointed out  in Section 3.

\section{Harmonic oscillator in twisted Moyal space}
As a prelude to the construction of a matrix basis appropriate for this study,
let us set up main algebraic relations pertaining to  twisted noncommutative coordinate
 transformations.
\subsection{Useful relations}
We consider the following infinitesimal affine transformation
\begin{eqnarray}
e_{a}^{\mu}(x)=\delta_{a}^{\mu}+\omega_{ab}^{\mu}x^{b},\quad \omega_{ab}^{\mu}=
:-\omega_{b a}^{\mu},\mbox{ and } |\omega^{\mu}|<<1.
\end{eqnarray}
In the sequel, we  restrict the discussion  to $D=2$, where
 $e_{a}^{\mu}$ and $\Theta^{ab}$ can be expressed as follows:
\begin{eqnarray}
(e)_{a}^{\mu}=\Big(\begin{array}{cc}
1+\omega_{12}^{1}x^{2}&\omega_{12}^{2}x^{2}\\
-\omega_{12}^{1}x^{1}&1-\omega_{12}^{2}x^{1}
\end{array}\Big) \quad \mbox{ and }\quad (\Theta)^{ab}=\Big(\begin{array}{cc}
0&\theta\\
-\theta&0
\end{array}\Big)=\theta (J)^{ab}
 \end{eqnarray}
where $J^{12}=-J^{21}=1,\,\, J^{11}=J^{22}=0$. At the first order
of expansion,
\begin{eqnarray}
e^{-1}=:det(e_{a}^{\mu})=1+\omega_{12}^{1}x^{2}-\omega_{12}^{2}x^{1}\\
 e=:det(e_{\mu}^{a})= 1-\omega_{12}^{1}x^{2}+\omega_{12}^{2}x^{1}.
\end{eqnarray}
 The $\star-$product of two Schwartz functions on
$\mathbb{R}^{2}_{\Theta}$   can be written under the form
\begin{eqnarray}\label{prod}
(f\star g)(x)={\rm m}\Big[\exp\Big(\frac{i}{2}\theta
e^{-1}J^{\mu\nu}\partial_\mu\otimes\partial_\nu\Big)(f\otimes
g)(x)\Big]
\end{eqnarray}
where $\mu,\nu=1,2$  and $\partial_{\mu}=:\frac{\partial}{\partial
x^{\mu}}$ .  Using the twisted star product (\ref{prod}) one can
see that
\begin{eqnarray}
e^{ikx}\star e^{iqx}=e^{i(k+q)x}e^{-\frac{i}{2}\theta e^{-1}kJq}.
\end{eqnarray}
The Fourier transform of $f , g\in
\mathcal{S}(\mathbb{R}_{\Theta}^{2})$   can be written as
\begin{eqnarray}
\tilde{f}(k)=\int d^2x \,\,e^{-ikx}f(x),\,\,\,\,\tilde{g}(q)=\int
d^2x \,\,e^{-iqx}g(x)
\end{eqnarray}
with the functions inverse transform given by
\begin{eqnarray}
f(x)=\frac{1}{(2\pi)^2}\int d^2k
\,\,e^{ikx}\tilde{f}(k),\,\,\,\,g(x)=\frac{1}{(2\pi)^2}\int d^2q
\,\,e^{iqx}\tilde{g}(q).
\end{eqnarray}
We can then redefine the twisted  star product of two Schwartz
functions $f , g$ as:
\begin{eqnarray}
(f\star g)(x)&=&\frac{1}{(2\pi)^4}\int
d^2kd^2q\,\,\tilde{f}(k)\tilde{g}(q)e^{ikx}\star e^{iqx}\cr
 &=& \frac{1}{(2\pi)^4}\int\,d^2kd^2q\,\,\tilde{f}(k)\tilde{g}(q)e^{i(k+q)x}e^{-\frac{i}{2}\theta
e^{-1}kJq}\cr
&=&\frac{1}{(2\pi)^4}\int\,d^2kd^2q\int\,d^2yd^2z\,\, f(y)g(z)\cr
&&\times e^{ik(x-y-\frac{1}{2}\theta e^{-1}Jq)}e^{iq(x-z)}
\end{eqnarray}
Using the identity
\begin{eqnarray}
\int\,d^2k\,\,e^{ik(x-y-\frac{1}{2}\theta
e^{-1}Jq)}=(2\pi)^{2}\delta^{(2)}(x-y-\frac{1}{2}\theta e^{-1}Jq)
\end{eqnarray}
and the variable change  $q$ to $q'=\frac{1}{2}\theta e^{-1}Jq$,
we arrive at   the adapted form for the proof of the next
Proposition \ref{p1.1}:
\begin{eqnarray}\label{18}
(f\star g)(x)&=&\Big(\frac{e}{\pi\theta}\Big)^2\int\,d^2yd^2z\,\,
f(y)g(z)e^{-\frac{2ei}{\theta }(x-y)J(x-z)}\cr
 &=&\Big(\frac{e}{\pi\theta}\Big)^2\int\,d^2yd^2z\,\,
f(x-y)g(x-z)e^{\frac{-2ei}{\theta}yJz}\cr
&=&\int\,d^2z\frac{d^2t}{(2\pi)^{2}}\,\,f(x-\frac{1}{2}\theta
e^{-1} t)g(x-z)e^{-itJz}.
\end{eqnarray}
\begin{proposition}\label{p1.1}
 If $f$ and $g$ are two Schwartz functions on
 $\mathbb{R}^{2}_{\Theta}$,
then $f\star g$ is also a Schwartz function on $\mathbb{R}^{2}_{\Theta}$.
\end{proposition}
{\bf Proof}: It is immediate by induction on the formulas
(\ref{leib}) and (\ref{18}) using integration by parts.
 $\square$\\
The tensor  $\widetilde{\Theta}^{\mu\nu}$ can be  explicit  as
\begin{eqnarray}
(\widetilde{\Theta})^{\mu\nu}=(\Theta)^{\mu\nu}-(\Theta^{a[\mu
}\omega_{ab}^{\nu]})x^b =\Big(\begin{array}{cc}
0& \theta e^{-1}\\
-\theta e^{-1}&0
\end{array}\Big).
\end{eqnarray}
 The twisted Moyal product of fields
generates some basic properties like the Jacobi identity
\begin{eqnarray}
 [x^\mu,[x^\nu, x^{\rho}]_{\star}]_{\star}+[x^\rho,[x^\mu, x^\nu]_{\star}]_{\star}
 +[x^\nu,[x^\rho, x^\mu]_{\star}]_{\star}=\Theta^{b\mu}\Theta^{d[\nu}\omega_{bd}^{\rho]}=0
\end{eqnarray}
conferring a Lie algebra structure to the defined twisted Moyal
space, and
\begin{eqnarray}\label{f}
x^{\mu}\star f=x^{\mu}f+\frac{i}{2}\Theta^{ab}e_{a}^{\mu}e_{b}^{\rho}\partial_{\rho}f
\;\mbox{ and }\,
f\star x^{\mu}=x^{\mu}f-\frac{i}{2}\Theta^{ab}e_{a}^{\mu}e_{b}^{\rho}\partial_{\rho}f.
\end{eqnarray}
The star brackets (anticommutator and commutator) of $x^{\mu}$ and
$f$ can be immediately deduced as follows:
$\{x^{\mu}, f\}_{\star}=2x^{\mu}f,\quad [x^{\mu}, f]_{\star}=i\Theta^{ab}e_{a}^{\mu}e_{b}^{\rho}\partial_{\rho}f$.
The relations (\ref{f}) can be detailed for $x^{\mu},\; \mu=1, 2$
as:
\begin{eqnarray}\label{11}
x^{1}\star f=x^{1}f+\frac{i}{2}\theta e^{-1}\partial_{2}f\quad\quad
f\star x^{1}=x^{1}f-\frac{i}{2}\theta e^{-1}\partial_{2}f
\end{eqnarray}
\begin{eqnarray}\label{12}
x^{2}\star f=x^{2}f-\frac{i}{2}\theta e^{-1}\partial_{1}f\quad\quad
f\star x^{2}=x^{2}f+\frac{i}{2}\theta e^{-1}\partial_{1}f
\end{eqnarray}
giving rise to the  creation and annihilation functions
\begin{eqnarray}
a=\frac{x^1 +ix^2}{\sqrt{2}}\quad\quad\quad \bar{a}=\frac{x^1-ix^2}{\sqrt{2}}
\end{eqnarray}
 with the commutation relation
$[a,\bar{a}]_{\star}=\theta e^{-1}$.
It then becomes a matter of algebra to use the  transformations of the vector fields $\partial_{1}$ and $\partial_{2}$
into $\partial_{a}=:\frac{\partial}{\partial a}$ and $\partial_{\bar{a}}=:\frac{\partial}{\partial \bar{a}}$ and vice-versa to infer
$e^{-1}=1-a\omega-\bar{a}\bar{\omega}$ and  $e= 1+a\omega+\bar{a}\bar{\omega}$,
where
\begin{eqnarray}
\omega=:\frac{\omega_{12}^{2}+i\omega_{12}^{1}}{\sqrt{2}}\quad\mbox{ and }\quad \bar{\omega}=:
\frac{\omega_{12}^{2}-i\omega_{12}^{1}}{\sqrt{2}}
\end{eqnarray}
leading to useful relations
\begin{eqnarray}
\frac{\partial e^{-1}}{\partial a}=-\omega,\quad \frac{\partial e^{-1}}{\partial \bar{a}}=-\bar{\omega}\mbox{ and for } k\in \mathbb{Z},\quad \omega e^{k}=\omega,
\quad \bar{\omega}e^{k}=\bar{\omega}.
\end{eqnarray}
Expressing the twisted $\star-$product (\ref{prod}) in terms of vectors fields $\partial_{a}$ and $\partial_{\bar{a}}$  as
\begin{eqnarray}\label{starstar}
(f\star g)(a,\bar{a})&=&{\rm m}\Big[\sum_{n=0}^{\infty}\sum_{k=0}^{n}
\frac{(-1)^{n-k}}{k!(n-k)!}(\frac{1}{2}\theta e^{-1})^{n}\cr
&&\times (\partial_{a}\otimes\partial_{\bar{a}})^{k}(\partial_{\bar{a}}\otimes\partial_{a})^{n-k}(f\otimes g)(a,\bar{a})\Big]
\end{eqnarray}
and using equations (\ref{11}) and (\ref{12}) (or independently (\ref{starstar})) yield
\begin{eqnarray}
a\star f=\Big(a+\frac{\theta e^{-1}}{2}\frac{\partial}{\partial \bar{a}}\Big)f \quad
\quad \bar{a}\star f=\Big(\bar{a}-\frac{\theta e^{-1}}{2}\frac{\partial}{\partial a}\Big)f
\end{eqnarray}
\begin{eqnarray}
f\star a=\Big(a-\frac{\theta e^{-1}}{2}\frac{\partial}{\partial \bar{a}}\Big)f\quad\quad f\star \bar{a}=\Big(\bar{a}+\frac{\theta e^{-1}}{2}\frac{\partial}{\partial a}\Big)f.
\end{eqnarray}
Provided the above definitions, we can now introduce the notions of right and left harmonic oscillator states denoted by
 $f_{m0}^{R}$ and $f_{0n}^{L}$, respectively.
\subsection{The right and left states}
Let $f_{00}^{R}\in L^{2}(\mathbb{R}_{\Theta}^{2})$  be the ho ``{\it right}'' fundamental state  such that
\begin{eqnarray}
a\star f_{00}^{R}=:0\quad\mbox{ with }\quad f_{00}^{R}=2e^{-\frac{2a\bar{a}}{\theta e^{-1}}(1-\frac{1}{2}\bar{a}\bar{\omega})}.
\end{eqnarray}
Then, $f_{00}^{R}$ solves the eigenvalue problem
$H\star f_{00}^{R}=\mathcal{E}_{00}^{R}f_{00}^{R}$
with  the corresponding right fundamental eigenvalue
$\mathcal{E}_{00}^{R}=\frac{\theta}{2}(1-2\bar{a}\bar{\omega})$ of the self-adjoint unbounded twisted ho Hamiltonian operator
\begin{eqnarray}
H\star(.)&=&:\bar{a}a\star(.)=\left[\bar{a}\star a+\frac{\theta e^{-1}}{2}\right]\star(.)=\left[a\star \bar{a}-\frac{\theta e^{-1}}{2}\right]\star(.)\cr
& = &{1\over 2}\left[(x^1)^2+(x^2)^2 + \left(i\theta e^{-1}x^1-{\theta^2 \over 4}\omega_{12}^1\right)\partial_2\right.\cr
&-&\left.\left(i\theta e^{-1}x^2-{\theta^2 \over 4}\omega_{12}^2\right)\partial_1
- {\theta^2 \over 4}e^{-2}(\partial_1^2 + \partial_2^2)\right]\equiv {1\over 2}\mu_1
\end{eqnarray}
defined in the domain
\begin{eqnarray}
 \mathcal{D}(H\star)= \left\{f\in L^{2}(\mathbb{R}_{\Theta}^{2})\mid f, f_{x^1}, f_{x^2} \in \mathcal{AC}_{loc}(\mathbb{R}_{\Theta}^{2});
{\mu_1 \over 2}f\in  L^{2}(\mathbb{R}_{\Theta}^{2})\right\}.
\end{eqnarray}
 $\mathcal{AC}_{loc}(\mathbb{R}_{\Theta}^{2})$
 denotes the set of the locally absolutely continuous functions on $\mathbb{R}_{\Theta}^{2}$. Similarly,
the fundamental left state $f_{00}^{L}\in L^{2}(\mathbb{R}_{\Theta}^{2})$  defined such that
\begin{eqnarray}
 f_{00}^{L}\star \bar{a}=:0\quad\mbox{ with }\quad f_{00}^{L}=2e^{-\frac{2a\bar{a}}{\theta e^{-1}}(1-\frac{1}{2}a\omega)}
\end{eqnarray}
solves the eigenvalue problem
$f_{00}^{L}\star H=\mathcal{E}_{00}^{L}f_{00}^{L}$
with the left fundamental eigenvalue $\mathcal{E}_{00}^{L}= \frac{\theta}{2}(1-2a\omega)$
 of the self-adjoint unbounded twisted ho Hamiltonian operator
\begin{eqnarray}
(.)\star H&=&:(.)\star\bar{a}a=(.)\star\left[\bar{a}\star a+\frac{\theta e^{-1}}{2}\right]=(.)\star\left[a\star \bar{a}-\frac{\theta e^{-1}}{2}\right]\cr
& = &{1\over 2}\left[(x^1)^2+(x^2)^2 - \left(i\theta e^{-1}x^1-{\theta^2 \over 4}\omega_{12}^1\right)\partial_2\right.\cr
&+&\left.\left(i\theta e^{-1}x^2-{\theta^2 \over 4}\omega_{12}^2\right)\partial_1
- {\theta^2 \over 4}e^{-2}(\partial_1^2 + \partial_2^2)\right]\equiv {1\over 2}\mu_2
\end{eqnarray}
defined in the domain
\begin{eqnarray}
 \mathcal{D}(\star H)= \left\{f\in L^{2}(\mathbb{R}_{\Theta}^{2})\mid f, f_{x^1}, f_{x^2} \in \mathcal{AC}_{loc}(\mathbb{R}_{\Theta}^{2});
{\mu_2 \over 2}f\in  L^{2}(\mathbb{R}_{\Theta}^{2})\right\}.
\end{eqnarray}
Then, the other states  follow from the next statement.
\begin{proposition}\label{p1}
The vectors $f_{m0}^{R}\in L^{2}(\mathbb{R}^{2}_{\Theta})$  given for any $m\in\mathbb{N}$ by
 \begin{eqnarray}\label{ls}
f_{m0}^{R}=\frac{1}{\sqrt{m!\theta^{m}}}\Big[2^{m}\bar{a}^{m}
(1+\frac{ma\omega}{2}-\frac{m\bar{a}\bar{\omega}}{4})-\frac{U_{m}\theta\omega\bar{a}^{m-1}}{2}\Big] f_{00}^{R}
\end{eqnarray}
solve the  eigenvalue problem $H\star f_{m0}^{R}= \mathcal{E}_{m0}^{R} f_{m0}^{R}$
with
\begin{eqnarray}
\mathcal{E}_{m0}^{R}=\frac{\theta}{2}\Big[2m+1-ma\omega -(3m+2)\bar{a}\bar{\omega}
-\frac{m^2 \theta\omega}{4\bar{a}}+\frac{\theta\omega U_{m}}{2^{m}\bar{a}}\Big],\,\, m\in \mathbb{N}
\end{eqnarray}
where
\begin{eqnarray}
 U_{m}=
(m-1)2^{m-2}+\sum_{k=0}^{m-3}(k+1)2^{k+1},\quad m\geq 3,\quad\quad
U_{i\leq 1}=0,\, U_{2}=1.\end{eqnarray}
\end{proposition}
{\bf Proof}: The results are immediate by induction, performing similar analysis as
 in \cite{Gracia-Bondia} to  construct
the right states $f^{R}_{m0}$  such that
$\bar{a}\star f^{R}_{m0}=\sqrt{\theta (m+1)}f^{R}_{m+1,0}$.
 $\square$\\
Similarly, the study of
the ho left states  provides  the following result.
\begin{proposition}
The vectors $f_{0n}^{L}\in L^{2}(\mathbb{R}^{2}_{\Theta})$  given for any $n\in \mathbb{N}$ by
\begin{eqnarray}
 f_{0n}^{L}=\frac{1}{\sqrt{n!\theta^{n}}}\Big[2^{n}a^{n}(1+\frac{n\bar{a}\bar{\omega}}{2}-\frac{na\omega}{4})
-\frac{U_{n}\theta\bar{\omega}a^{n-1}}{2}\Big]f_{00}^{L}
\end{eqnarray}
solve the  eigenvalue problem  $f_{0n}^{L}\star H= \mathcal{E}_{0n}^{L} f_{0n}^{L}$
with
\begin{eqnarray}
\mathcal{E}_{0n}^{L}=\frac{\theta}{2}\Big[2n+1-n\bar{a}\bar{\omega} -(3n+2)a\omega
-\frac{n^2 \theta\bar{\omega}}{4a}+\frac{\theta\bar{\omega}U_{n}}{2^{n}a}\Big],\,\, n\in\mathbb{N}.
\end{eqnarray}
\end{proposition}
{\bf Proof}: It uses  the same procedure as  previously, but with the construction of
the left states $f^{L}_{0n}$ such that
$f^{L}_{0n}\star a=\sqrt{\theta (n+1)}f^{L}_{0,n+1}$.
$\square$\\
Besides $\lim_{\omega,\bar{\omega}\rightarrow 0}\mathcal{E}_{m0}^{R}=\theta\Big(m+\frac{1}{2}\Big) \,\mbox{ and }\,
 \lim_{\omega,\bar{\omega}\rightarrow 0}\mathcal{E}_{0n}^{L}=\theta\Big(n+\frac{1}{2}\Big)$
corresponding to the usual Moyal $\star-$product spectrum of the harmonic oscillator Hamiltonian $H$.

All these results show that the ho right and left states as well as their respective energy
spectrum are expressible in terms
 of the space deformation constant $\theta$ and of an additional piece inherent to the nature
 of the induced infinitesimal
transformation through the parameter $\omega$ and its conjugate.
Besides, a noteworthy feature of these states is the following.
\begin{proposition}\label{2.3}
 The right and left  fundamental states defined by $f_{00}^{(m)
 R}=:a^{m+1}\star f_{m0}^{R}$ and $f_{00}^{(n)L}=:f_{0n}^{L}\star \bar{a}^{n+1}$
 are given by the following expressions:
\begin{eqnarray}
  f_{00}^{(m)R}&=&-\frac{\sqrt{m!\theta^{m+2}}}{8}\sum_{j=1}^{m}\frac{(m+4j+1)}{(m-j)!}
\bar{\omega} f_{00}^{R}\,\, \mbox{ and }\\
f_{00}^{(n)L}&=& f_{00}^{(n)R}(\bar{a}\leftrightarrow a,\bar{\omega}\leftrightarrow\omega)\cr
&=&-\frac{\sqrt{n!\theta^{n+2}}}{8}\sum_{j=1}^{n}\frac{(n+4j+1)}{(n-j)!}\omega f_{00}^{L}
\end{eqnarray}
which, in the usually Moyal product case, are reduced to $0$.
Besides, the twisted  harmonic oscillator states $f_{m0}^{R}$ and $f_{m0}^{L}$ are degenerate with respect to the rules
\begin{eqnarray}
a^{m+2}\star f_{m0}^{R}=0 \;\;\mbox{and}\;\;  f_{0m}^{L}\star \bar{a}^{m+2}= 0\,\,\; \forall \; m\ge 1.
\end{eqnarray}
\end{proposition}
The proof is straightforward. See appendix A.
\begin{remark}
\begin{enumerate}
\item $f_{00}^{R}$ and $f_{00}^{L}$ are the twisted fundamental states restoring, in the limit
of ordinary Moyal space, the fundamental state given by $2e^{-\frac{2a\bar{a}}{\theta}}$.
For the analysis purpose, we  call $f_{00}^{R}$ and $f_{00}^{L}$  the normal twisted fundamental states.
\item The states $f_{m0}^{R}$ correspond to twisted right $m+1$ particles states, reducing, in the usual case,
to right $m$ particles states, while the states $f_{0n}^{L}$  represent  the twisted left $n+1$ particles states.
\item There are an infinite number of twisted right $m-k$ particles states and
  an infinite number of twisted left $n-k$ particles states given by
$a^{k+1}\star f_{m0}^{R}$  and $f_{0n}^{L}\star\bar{a}^{k+1}$, respectively.
\end{enumerate}
\end{remark}
\subsection{Matrix basis of the theory}
The usual construction of a matrix basis \cite{Gracia-Bondia} exploits
 the $\star-$multiplication of $f_{m0}^{R}$ with $f_{0n}^{L}$, i.e.
\begin{eqnarray}
 L^{2}(\mathbb{R}_{\Theta}^{2}) \ni b^{(2)}_{mn}&=&:\chi(\omega, \bar{\omega},\theta, m,n) f_{m0}^{R}\star f_{0n}^{L}
\nonumber\\
&=&\chi(\omega, \bar{\omega},\theta,m,n)\frac{\bar{a}^{m}\star f_{00}^{R}\star f_{00}^{L}\star a^{n}}{\sqrt{m!n!\theta^{m+n}}}.
\end{eqnarray}
 Without loss of generality, we set the normalization constant 
$\chi(\omega, \bar{\omega},\theta,m,n)=:1$
by  convention. The  corresponding eigenvalue problems are given by
\begin{eqnarray}
H\star b^{(2)}_{mn}=\mathcal{E}_{m0}^{R}b^{(2)}_{mn}\quad\mbox{ and }\quad b^{(2)}_{mn}\star H= \mathcal{E}_{0n}^{L}b^{(2)}_{mn}
\end{eqnarray}
while the $\star-$actions of the annihilation and creation  functions $a$ and $\bar a$ are reproduced as follows:
$\bar{a}\star b^{(2)}_{mn}=\sqrt{\theta(m+1)}b^{(2)}_{m+1,n}$ and $b^{(2)}_{mn}\star a=\sqrt{\theta(n+1)}b^{(2)}_{m,n+1}$,
with the basis fundamental state
$ b^{(2)}_{00}= f_{00}^{R}\star f_{00}^{L}$
 satisfying the expected requirements
$a\star b^{(2)}_{00}=0,\quad b^{(2)}_{00}\star \bar{a}=0$.
 Given the  $(1,1)-$particles states defined by
$L^{2}(\mathbb{R}_{\Theta}^{2}) \ni \Lambda_{mn}^{1,1}=:a^{m}\star b_{mn}^{(2)}\star \bar{a}^{n}$,
their twisted spectrums can be computed from the eigenvalue problems
$H\star \Lambda_{mn}^{1,1}=\mathcal{E}_{\Lambda_{m0}^{1,1}}^R \Lambda_{mn}^{1,1}\,\mbox{ and }\,
\Lambda_{mn}^{1,1}\star H=\mathcal{E}_{\Lambda_{0n}^{1,1}}^{L}\Lambda_{mn}^{1,1}$
to get, depending on the right and left Hamiltonian $\star-$actions,
\begin{eqnarray}\label{qu3}
 \mathcal{E}_{\Lambda_{m0}^{1,1}}^{R}&=&\frac{\theta}{2}\Big[1-\frac{\bar{a}\bar{\omega}}{2}
\Big(\sum_{j=1}^{m}\frac{m+4j+1}{(m-j)!}+4\Big)\Big],\, m> 0 \\
\label{qu4} \mathcal{E}_{\Lambda_{0n}^{1,1}}^{L}&=&
\mathcal{E}_{\Lambda_{n0}^{1,1}}^{R}(\bar{a}\leftrightarrow a,\bar{\omega}\leftrightarrow \omega)\cr
&=&\frac{\theta}{2}
\Big[1-\frac{a\omega}{2}\Big(\sum_{j=1}^{n}\frac{n+4j+1}{(n-j)!}+4\Big)\Big],\quad n> 0.
\end{eqnarray}
For $\omega=0$ and $\bar{\omega}=0$, these energies are reduced to the usual Moyal space matrix basis right and left
fundamental energies.
As needed, the Wick rotation can be  used to ensure the real value of the energy.
In the same vein, one can define the single twisted $(m-k+1)$ right particles states
 by $a^{k}\star f_{m0}^{R}=:\Lambda_{m0}^{m-k+1}\in L^{2}(\mathbb{R}_{\Theta}^{2})$
corresponding to the energy values obtained from the right Hamiltonian  $\star-$action  by
\begin{eqnarray}\label{qu5}
\mathcal{E}_{\Lambda_{m0}^{m-k+1}}^{R}&=&\frac{\theta}{2}\Big\{2m-2k+1-(m-k)a\omega\nonumber\\
&+&\frac{\bar{a}\bar{\omega}}{2}\Big[(m-k-1)(m-k)!\sum_{j=1}^{k}\frac{m+4j+1}{(m-j)!}\nonumber\\
&-&(m-k)(m+4k+6)-4\Big]-\frac{(m-k)(m-2k)\theta\omega}{4\bar{a}}\nonumber\\
&+&\frac{(m-k+1)(m-k)\theta\omega U_m}{m2^{m+1}\bar{a}}\Big\}\quad m\geq k> 0.
\end{eqnarray}
By analogy,  the single  twisted $(n-l+1)$ left particles states $f_{0n}^{L}\star \bar{a}^{l}=:\Lambda_{0n}^{n-l+1}\in L^{2}(\mathbb{R}_{\Theta}^{2})$
 are associated with the left action energy values
\begin{eqnarray}\label{qu6}
\mathcal{E}_{\Lambda_{0n}^{n-l+1}}^{L}&=&\mathcal{E}_{\Lambda_{n0}^{n-l+1}}^{R}\Big(\bar{\omega}\leftrightarrow\omega,\bar{a}\leftrightarrow a\Big)\cr
&=&\frac{\theta}{2}\Big\{2n-2l+1-(n-l)\bar{a}\bar{\omega}\nonumber\\
&+&\frac{a\omega}{2}\Big[(n-l-1)(n-l)!\sum_{j=1}^{l}\frac{n+4j+1}{(n-j)!}\nonumber\\
&-&(n-l)(n+4l+6)-4\Big]-\frac{(n-l)(n-2l)\theta\bar{\omega}}{4a}\nonumber\\
&+&\frac{(n-l+1)(n-l)\theta\bar{\omega}U_n}{n2^{n+1}a}\Big\},\quad  n\geq l> 0.
\end{eqnarray}
\begin{proposition}
The energy spectrums (\ref{qu5}) and  (\ref{qu6}) of the {\it mixed} twisted $(m-k+1)$
right and $(n-l+1)$ left particles states
$ L^{2}(\mathbb{R}_{\Theta}^{2})\ni \Lambda_{mn}^{m-k+1,n-l+1}=: a^{k}\star f_{m0}^{R}\star f_{0n}^{L}\star \bar{a}^{l}$
solve the following respective eigenvalue problems:
\begin{eqnarray}
 H\star \Lambda_{mn}^{m-k+1,n-l+1}= \mathcal{E}_{\Lambda_{m0}^{m-k+1}}^{R}\Lambda_{mn}^{m-k+1,n-l+1}\\
\Lambda_{mn}^{m-k+1,n-l+1}\star H= \mathcal{E}_{\Lambda_{0n}^{n-l+1}}^{L}\Lambda_{mn}^{m-k+1,n-l+1}.
\end{eqnarray}
\end{proposition}
We readily recover the spectrums (\ref{qu3}) and (\ref{qu4}) by
replacing $m=k$ and $n=l$ in the relations
 (\ref{qu5}) and  (\ref{qu6}).
Of course, in the limit regime, these energies also well reproduce   the
ordinary Moyal plane  $(m-k)$ right
and $(n-l)$ left  particles energies:
\begin{eqnarray}
 \lim_{\omega,\bar{\omega}\rightarrow 0}\mathcal{E}_{\Lambda_{m0}^{m-k+1}}^{R}=\frac{\theta}{2}\Big[2(m-k)+1\Big]
\\
\lim_{\omega,\bar{\omega}\rightarrow 0} \mathcal{E}_{\Lambda_{0n}^{n-l+1}}^{L}= \frac{\theta}{2}\Big[2(n-l)+1\Big]
\end{eqnarray}
respectively.
\section{Concluding remarks}
We have investigated the main properties of harmonic oscillator in the framework of a dynamical noncommutativity realized
through a twisted Moyal product. The construction of the appropriate matrix basis  has introduced
an $x$-dependence in the definition of the star product.
We have computed  the states and energies of the twisted harmonic oscillator. The degeneracy states and their energy
 have been explicitly
 derived. All  examined quantities easily acquire good  physical  properties
 when $\omega_{12}^{2}$ and $x^1$
are transformed  into $i\omega_{12}^{2}$ and $ix^1$, respectively.
Furthermore,  the ordinary Moyal space tools are well  recovered
as expected by setting $\omega=0$ and $\bar{\omega}=0$.

\section*{Acknowledgements}
This work is partially supported by the ICTP through the
OEA-ICMPA-Prj-15. The ICMPA is in partnership with the Daniel
Iagolnitzer Foundation (DIF), France.
\section*{Appendix A: Right and left $\star-$actions of the
 creation and annihilation functions onto the ho states}
In this part, we derive the right and left $\star-$actions of the
 creation and annihilation functions onto the ho states.
\begin{eqnarray}
a\star f_{m0}^{R}&=&\frac{m2^{m-1}\bar{a}^{m-1}}{\sqrt{m!\theta^{m-2}}}(1+\frac{m-2}{2}a\omega-\frac{m+5}{4}\bar{a}\bar{\omega})f_{00}^{R}\nonumber\\
&-&\frac{(m-1)\theta\omega U_{m}\bar{a}^{m-2}}{4\sqrt{m!\theta^{m-2}}}f_{00}^{R}\cr
a^{2}\star f_{m0}^{R}&=&\frac{m(m-1)2^{m-2}\bar{a}^{m-2}}{\sqrt{m!\theta^{m-4}}}\Big[1+\frac{m-4}{2}a\omega\nonumber\\
&-&\frac{(m+9)(m-1)+m+5}{4(m-1)}\bar{a}\bar{\omega}\Big]f_{00}^{R}\nonumber\\
&-&\frac{(m-1)(m-2)\theta\omega U_{m}\bar{a}^{m-3}}{8\sqrt{m!\theta^{m-4}}}f_{00}^{R}\cr
&&\vdots\nonumber\\
a^{k}\star f_{m0}^{R}&=&\frac{m(m-1)\cdots(m-k+1)2^{m-k}\bar{a}^{m-k}}{\sqrt{m!\theta^{m-2k}}}\Big[1+\frac{m-2k}{2}a\omega\nonumber\\
&-&\frac{\bar{a}\bar{\omega}}{4}(m-k)!\sum_{j=1}^{k}\frac{(m+4j+1)}{(m-j)!}\Big]f_{00}^{R}\nonumber\\
&-&\frac{(m-1)(m-2)\cdots (m-k)\theta\omega U_{m}\bar{a}^{m-k-1}}{2^{k+1}\sqrt{m!\theta^{m-2k}}}f_{00}^{R}\cr
&&\mbox{ where } k\leq m\cr
\vdots\nonumber\\
a^{m}\star f_{m0}^{R}&=&\frac{m!}{\sqrt{m!\theta^{-m}}}\Big[1-\frac{ma\omega}{2}-\frac{\bar{a}\bar{\omega}}{4}\sum_{j=1}^{m}
\frac{(m+4j+1)}{(m-j)!}\Big]f_{00}^{R}\cr
a^{m+1}\star f_{m0}^{R}&=&-\frac{\sqrt{m!\theta^{m+2}}\bar{\omega}}{8}\sum_{j=1}^{m}\frac{(m+4j+1)}{(m-j)!}f_{00}^{R}
\propto f_{00}^{R}\\
 a^{m+2}\star f_{m0}^{R}&=&0.
\end{eqnarray}
Similarly, $f_{0n}^{L}\star\bar{a} $,
$f_{0n}^{L}\star\bar{a}^2,\cdots f_{0n}^{L}\star\bar{a}^n,
f_{0n}^{L}\star\bar{a}^{n+1}$ can be computed as
\begin{eqnarray}
f_{0n}^{L}\star\bar{a}&=&\frac{n2^{n-1}a^{n-1}}{\sqrt{n!\theta^{n-2}}}(1+\frac{n-2}{2}
\bar{a}\bar{\omega}-\frac{n+5}{4}a\omega)f_{00}^{L}\nonumber\\
&-&\frac{(n-1)\theta\bar{\omega} U_{n}a^{n-2}}{4\sqrt{n!\theta^{n-2}}}f_{00}^{L}\cr
f_{0n}^{L}\star
\bar{a}^{2}&=&\frac{n(n-1)2^{n-2}a^{n-2}}{\sqrt{n!\theta^{n-4}}}
\Big[1+\frac{n-4}{2}\bar{a}\bar{\omega}\nonumber\\
&-&\frac{(n+9)(n-1)+n+5}{4(n-1)}a\omega\Big]f_{00}^{L}\nonumber\\
&-&\frac{(n-1)(n-2)\theta\bar{\omega} U_{n}a^{n-3}}{8\sqrt{n!\theta^{n-4}}}f_{00}^{L}\cr
\vdots\nonumber\\
f_{0n}^{L}\star\bar{a}^{k}&=&\frac{n(n-1)\cdots(n-k+1)2^{n-k}a^{n-k}}{\sqrt{n!\theta^{n-2k}}}
\Big[1+\frac{n-2k}{2}\bar{a}\bar{\omega}\nonumber\\
&-&\frac{a\omega}{4}(n-k)!\sum_{j=1}^{k}\frac{(n+4j+1)}{(n-j)!}\Big]f_{00}^{L}\nonumber\\
&-&\frac{(n-1)(n-2)\cdots (n-k)\theta\bar{\omega}
U_{n}a^{n-k-1}}{2^{k+1} \sqrt{n!\theta^{n-2k}}}f_{00}^{L};\cr
&&\mbox{ where } k\leq n\cr
\vdots\nonumber\\
f_{0n}^{L}\star\bar{a}^{n}&=&\frac{n!}{\sqrt{n!\theta^{-n}}}\Big[1-\frac{n\bar{a}
\bar{\omega}}{2}-\frac{a\omega}{4}\sum_{j=1}^{n}\frac{(n+4j+1)}{(n-j)!}\Big]f_{00}^{L}\cr
f_{0n}^{L}\star \bar{a}^{n+1}&=&-\frac{\sqrt{n!\theta^{n+2}}\omega}{8}\sum_{j=1}^{n}\frac{(n+4j+1)}{(n-j)!}f_{00}^{L}\propto f_{00}^{L}\\
  f_{0n}^{L}\star \bar{a}^{n+2}&=&0.
\end{eqnarray}

\section*{Appendix B: Useful identities}
\begin{eqnarray}
\partial_{a}^{k}f_{00}^{R}=\Big[k(k-1)\omega\Big(-\frac{2\bar{a}}{\theta}\Big)^{k-1}+\Big(-\frac{2\bar{a}}{\theta e^{-1}}\Big)^{k}(1+ka\omega-\frac{k\bar{a}\bar{\omega}}{2})\Big]f_{00}^{R},
\end{eqnarray}
\begin{eqnarray}
\partial_{\bar{a}}^{k}f_{00}^{R}=\Big[\frac{k(k-1)}{2}\bar{\omega}\Big(-\frac{2a}{\theta}\Big)^{k-1}+\Big(-\frac{2a}{\theta e^{-1}}\Big)^k\Big] f_{00}^{R}
\end{eqnarray}
\begin{eqnarray}
\partial_{\bar{a}}^{k}f_{00}^{L}=\Big[k(k-1)\bar{\omega}\Big(-\frac{2a}{\theta}\Big)^{k-1}+\Big(-\frac{2a}{\theta e^{-1}}\Big)^{k}(1+k\bar{a}\bar{\omega}-\frac{ka\omega}{2})\Big]f_{00}^{L},
\end{eqnarray}
\begin{eqnarray}
\partial_{a}^{k}f_{00}^{L}=\Big[\frac{k(k-1)}{2}\omega\Big(-\frac{2\bar{a}}{\theta}\Big)^{k-1}+\Big(-\frac{2\bar{a}}{\theta e^{-1}}\Big)^k\Big] f_{00}^{L}.
\end{eqnarray}

\begin{eqnarray}
\partial_{a}^{k}\Big(-\frac{2a}{\theta e^{-1}}\Big)^{l}&=&kl\omega\frac{l!}{(l-k+1)!}\Big(-\frac{2}{\theta}\Big)^{k-1}\Big(-\frac{2a}{\theta}\Big)^{l-k+1}\cr
&&+\frac{l!}{(l-k)!}\Big(-\frac{2}{\theta}\Big)^{k}\Big(-\frac{2a}{\theta}\Big)^{l-k}
e^{l}
\end{eqnarray}

\begin{eqnarray}
\partial_{\bar{a}}^{k}\Big(-\frac{2\bar{a}}{\theta e^{-1}}\Big)^{l}&=&kl\bar{\omega}\frac{l!}{(l-k+1)!}\Big(-\frac{2}{\theta}\Big)^{k-1}\Big(-\frac{2\bar{a}}{\theta}\Big)^{l-k+1}\cr
&&+\frac{l!}{(l-k)!}\Big(-\frac{2}{\theta}\Big)^{k}\Big(-\frac{2\bar{a}}{\theta}\Big)^{l-k}
e^{l}.
\end{eqnarray}

\end{document}